\documentclass[%
 reprint,
 amsmath,amssymb,
 aps,
]{revtex4-2}

\usepackage{graphicx}
\usepackage{dcolumn}
\usepackage{bm}
\begin{document}

\title{ Ultrathin carpet cloak enabled by  infinitely anisotropic medium } 

\author{Mohammad Hosein Fakheri}
\author{Ali Abdolali}
\email{Abdolali@iust.ac.ir}
\affiliation{
	Applied Electromagnetic Laboratory, School of Electrical Engineering, Iran University of Science and Technology, Tehran, 1684613114, Iran.}

\begin{abstract}
Thanks to the pioneering studies conducted on the fields of transformation optics (TO) and metasurfaces, many unprecedented devices such as invisibility cloaks have been recently realized. However, each of these methods has some drawbacks limiting the applicability of the designed devices for real-life scenarios. For instance, TO studies lead to bulky coating layer with the thickness that is comparable to, or even larger than the dimension of the concealed object. In this paper, based on the coordinate transformation, an ultrathin carpet cloak is proposed to hide objects with arbitrary shape and size using a thin  anisotropic material, called as infinitely anisotropic medium (IAM). It is shown that unlike the previous metasurface-based carpet cloaks,  the proposed IAM hides objects from all viewing incident angles while it is extremely thin compared with the object dimensions. This material also circumvents the conventional transformation optics’ complexities and could be easily implemented in practical scenarios. To demonstrate the capability of the proposed carpet cloak, several full-wave simulations are carried out.  Finally, as a proof of concept, the IAM is implemented based on the effective medium theory which exhibits good agreement with the results obtained from the theoretical investigations. The introduced material not only constitutes a significant step towards the invisibility cloak but also can greatly promote the practical application of the other TO-based devices.

\end{abstract}

\maketitle

An invisibility cloak that can conceal objects from the electromagnetic (EM) wave  has attracted considerable attentions for centuries, but its realization remained unattainable until the advent of metamaterials. To design such an unprecedented device, various techniques such as, scattering cancelation \cite{Alu} and transformation optics \cite{Pendry}, have been proposed. However, the method of scattering cancelation is limited to small hidden objects in comparison to the operative wavelength \cite{Alu}. From the viewpoint of TO and based on the form invariance of Maxwell’s equations, an initial coordinate system (\emph{i.e.,}  virtual domain) is transformed to its counterpart in another arbitrary coordinate system (\emph{i.e.,} physical domain), which results in a direct link between the material properties and metric tensor of the transformed space having the desired EM functionalities \cite{kwon, Abdolali, Fakheri, Lai}. However, the  materials obtained through this approach are inhomogeneous and anisotropic, which are difficult to be implemented. To obviate the anisotropy issue, the method of quasi-conformal mapping is utilized \cite{Li}. Nevertheless, even by performing such a simplifying assumption, this cloak causes a lateral shift of the scattered wave, whose value is comparable to the height of the cloaked object, making the object detectable \cite{Zhang}. Moreover, the mentioned techniques  generally yield a bulky coating later with a thickness comparable to, or even larger than the dimension of the concealed object. 
As the 2D version of metamaterials, metasurfaces can control the EM waves via imparting abrupt phase changes determined by subwavelength elements at the interface and have overcame the certain drawbacks of bulky metamaterials \cite{Yu,Behroozinia,Rajabalipanah}. These ultrathin structures have paved the way towards achieving carpet cloaks hiding objects on the ground plane\cite{Jing,Huang}. Despite claims about wide-angle metasurface carpet cloak \cite{Jiang}, their functionality is restricted to a specific angular range. Indeed,  the cloaking performance is drastically deteriorated out of these incident wave angles. Consequently, for the above-mentioned reasons, one must ask the question of whether there is any alternative way to conceal an object with an ultrathin carpet cloak, regardless of the incident wave angle.

Here, we successfully design an ultrathin all-angle coating layer to conceal objects with arbitrary size and shape located on the ground, based on a special type of material named infinitely anisotropic medium (IAM). It will be demonstrated that the introduced IAM is independent of the concealed object geometry in such a way that if the object shape is changed, one can still use the same IAM. Several numerical simulations are carried out to verify the validity of the propounded invisibility cloak. It is observed that numerical results are in good agreement with the theoretical investigations indicating the validity of the presented approach. As a proof of concept, an ultrathin carpet cloak is realized through the effective medium theory (EMT) for which, the spiral resonator (SR)-meander line metamaterial blocks have been elaborately engineered to play the role of the required IAM. 

\begin{figure*}[!t]
	\centering
	\includegraphics[width=0.7\linewidth]{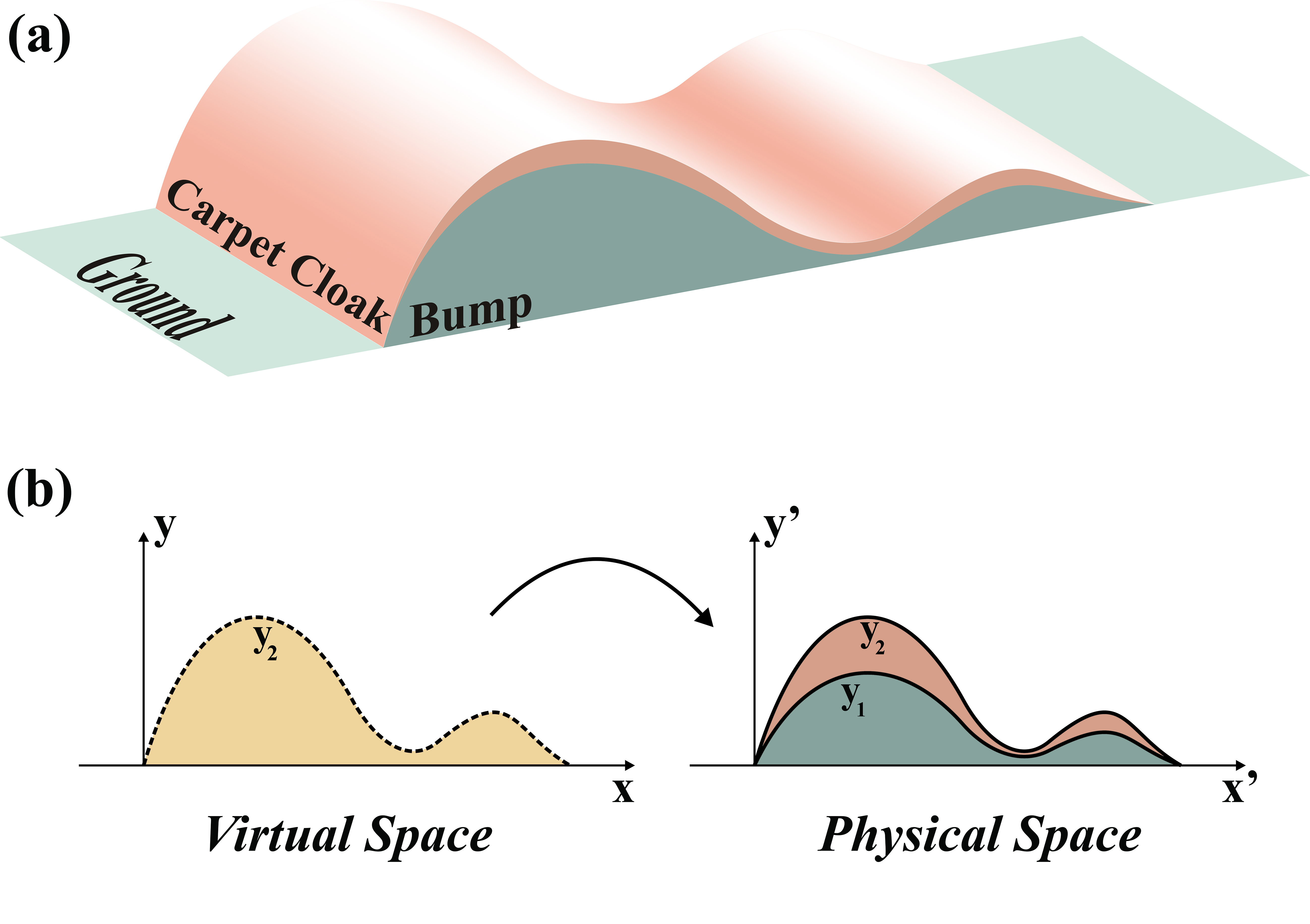}
	\caption{ (a) Schematic view of the proposed ultrathin carpet cloak for arbitrary shape object. (b) Designing a carpet cloak based on coordinate transformation. a cream region in the virtual space is mapped to a brown region in the physical space.}
	\label{fgr:fig1}
\end{figure*}
Fig. 1(a) shows the schematic view of the proposed carpet cloak that conceals the arbitrary shape object with an ultrathin coating layer. Let us start with a straightforward coordinate transformation in the general form. In order to conceal an arbitrary shape object with the curve of $y_1(x)$, the schematic of Fig. 1(b) is used as the space transformation. As can be seen, a cream region enclosed between the ground $\left( y=0 \right)$ and curve $y_2(x)$ in the virtual space is compressed into the brown region encircled by $y_1(x)$  and $y_2(x)$  in the physical space. Indeed, the ground is transformed to the concealed object boundary with the curve $y_1=f(x)$  while  the boundary $y_2=\Gamma f(x)$ is mapped to itself  keeping the exterior EM fields unchanged. 
 \par 
 
The transformation function of such a mapping could be written as $x'=x$, $y'=\alpha y+f(x)$ and $z'=z$
where $\alpha=\left( \Gamma -1 \right)/\Gamma $. According to the coordinate transformation theory, when a space $(x, y, z)$ is transformed into another space $(x',y',z')$
of different shape and size, the permittivity $\overline{\overline{{{\varepsilon }'}}}$ and permeability $\overline{\overline{{{\mu }'}}}$ values in the transformed space are given by $\overline{\overline{{{\varepsilon }'}}}={\Lambda \varepsilon {{\Lambda }^{T}}}/{\det \Lambda }\;$ and $\overline{\overline{{{\mu }'}}}={\Lambda \mu {{\Lambda }^{T}}}/{\det \Lambda }\;$, where $\varepsilon$ and $\mu$ refer to the permittivity and permeability in the original space and $\Lambda ={\partial \left( {x}',{y}',{z}' \right)}/{\partial \left( x,y,z \right)}\;$ indicates the Jacobian transformation tensor \cite{Pendry}. Consequently, the constitutive parameters of the proposed carpet cloak could be
expressed as

\begin{equation}
\frac{\overline{\overline{\varepsilon'}}}{\varepsilon}=\frac{\overline{\overline{\mu'}}}{\mu}=\frac{1}{\alpha}
\begin{bmatrix}
\ 1 &
f'(x)&
0 \\
f'(x) &

\alpha^2+(f'(x))^2 &
0\\
0 &
0 &
1
\end{bmatrix} 
\end{equation}
\par 
The obtained material of Eq.(1) is inhomogeneous and anisotropic with off-diagonal components of $f'(x)$, which may cause serious difficulties in its realization process. However, as we aim to design an ultrathin carpet cloak, we set $\Gamma \rightarrow 1$ which leads to $\alpha \rightarrow 0$. The diagonal form of Eq.(1) is obtained by rotating the matrix around its principle axis at the angle of $\varphi=tan^{-1}(f'(x))+\pi/2$ \cite{Barati}. Therefore, the diagonal form of Eq.(1) is expressed as

\begin{equation}
\frac{\overline{\overline{\varepsilon'_d}}}{\varepsilon}=\frac{\overline{\overline{\mu'_d}}}{\mu}=
\begin{bmatrix}
\ 0 &
0&
0 \\
0 &

\frac{1+(f'(x))^2}{\alpha} &
0\\
0 &
0 &
\frac{1}{\alpha}
\end{bmatrix} 
\overset{\alpha \to 0}{\mathop =}\,
\begin{bmatrix}
\ 0 &
0&
0 \\
0 &

\infty &
0\\
0 &
0 &
\infty
\end{bmatrix} 
\end{equation}
The material achieved by Eq. (2) is called an IAM with the main axis along the $x'$ direction. Therefore, by considering Eq.(2) and the rotation angle $\varphi$, one can conclude that an ideal ultrathin carpet cloak is composed of the IAM whose main axis is perpendicular to the object's surface. 

\begin{figure}[h]
	\centering
	\includegraphics[width=\linewidth]{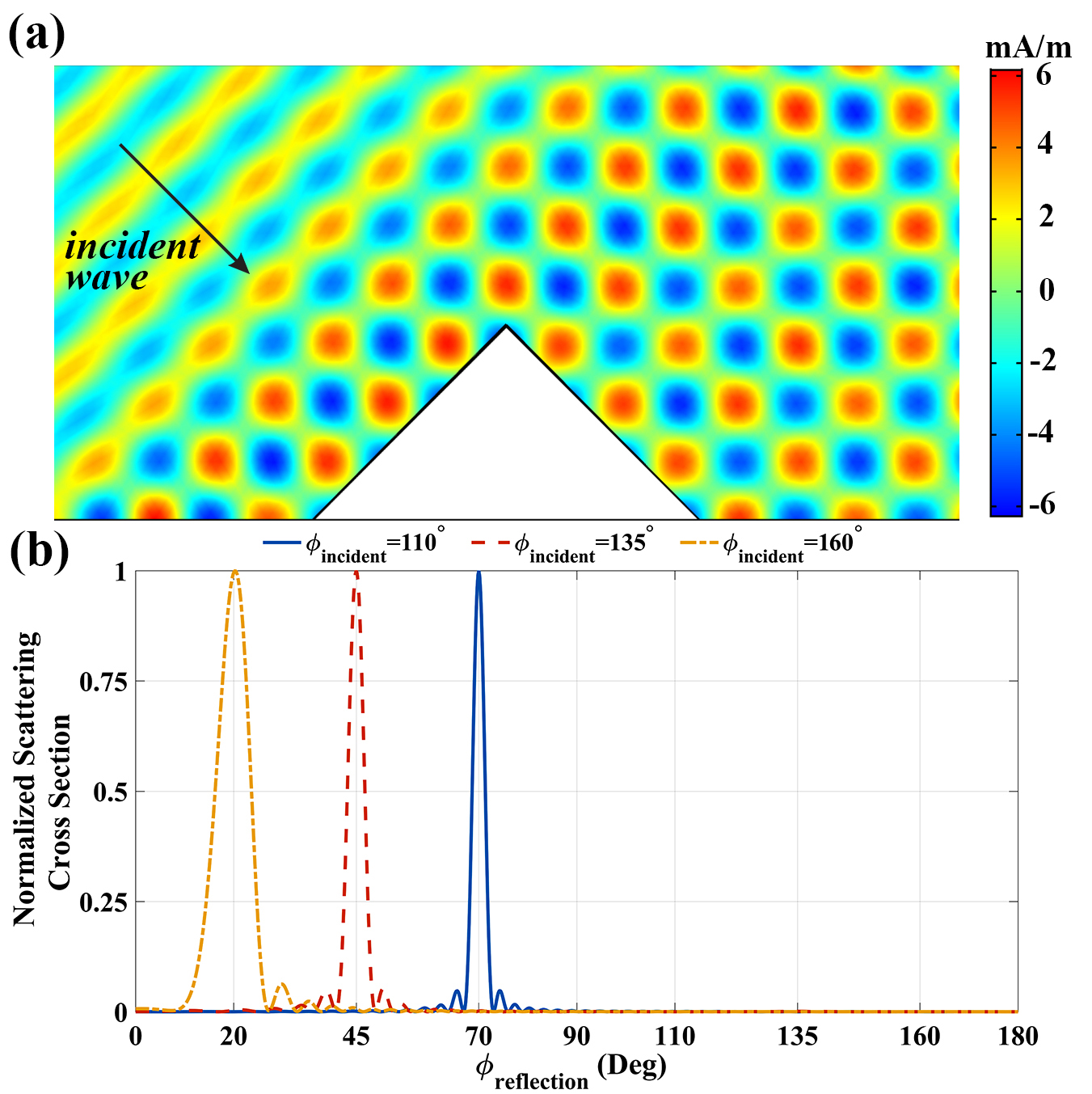}
	\caption{ Performance of the invisibility carpet cloak based on IAM for a triangle shape object with $\alpha=0.01$. (a) Magnetic field distribution at $\phi_{inc}=135^\circ$ . (b) Far-field patterns at $\phi_{inc}=110^\circ$, $\phi_{inc}=135^\circ$ and $\phi_{inc}=160^\circ$.}
	\label{fgr:fig2}
\end{figure}

In the following, we carry out full-wave simulations
using COMSOL Multiphysics finite element solver to demonstrate the functionality of ultrathin carpet cloaks to conceal arbitrary shape objects from all viewing angles. We assume the structure is illuminated by a TM-polarized ($H$ along the $z$ direction)  plane wave at $f$=3.5 GHz. The first example is dedicated to a
triangle object with the base and height parameters of $0.2m$ and $0.4m$, respectively. Since, selecting the exact value of 0 for $\alpha$ causes some errors in the simulation process, we set $\alpha=0.01$. Fig. 2(a) illustrates the near field
distributions of the magnetic field where the object is covered with the proposed IAM. As can be seen, under the incident wave angle of $\phi_{inc}=135^\circ$, the object becomes invisible, where the near field distribution mimics the specular reflection behavior of a flat ground plane. To further demonstrate the effectiveness of the proposed ultrathin cloak, the structure is examined under two other incident wave angles  $\phi_{inc}=110^\circ$ and $\phi_{inc}=160^\circ$. The scattering cross section of three mentioned cases are plotted in Fig. 2(b), where a single beam is noticed at the desired angle for each case. The difference of their beam width comes from the fact that  the more oblique the incident wave, the smaller the equivalent aperture size \cite{Balanis}. Consequently,the presented IAM can hide objects from all range of incident angles while having an extremely thin profile in comparison to the object size.\par

Moreover, for the objects with arbitrary shape that are
important in the practical scenarios, the previously reported approaches are not applicable due to the necessity to inhomogeneous and shape-dependent materials. Nevertheless, the material introduced in  Eq.(2), can mitigate most of these drawbacks. To this aim, an arbitrary shape object is coated by a thin IAM layer with $\alpha=0.03$. As illustrated in Fig. 3(a), the total magnetic field distribution of the concealed object is identical to that of the ground plane as if the object does not exist. Moreover, in contrast to the previous metasurface cloaks, when the angle of incidence  varies, the functionality  of our designed cloak remains unchanged, as validated with the scattering cross section in Fig. 3(b). In other words, the functionality of the cloak presented here is not restricted to a specific range of incident wave angles.

\begin{figure}[h]
	\centering
	\includegraphics[width=\linewidth]{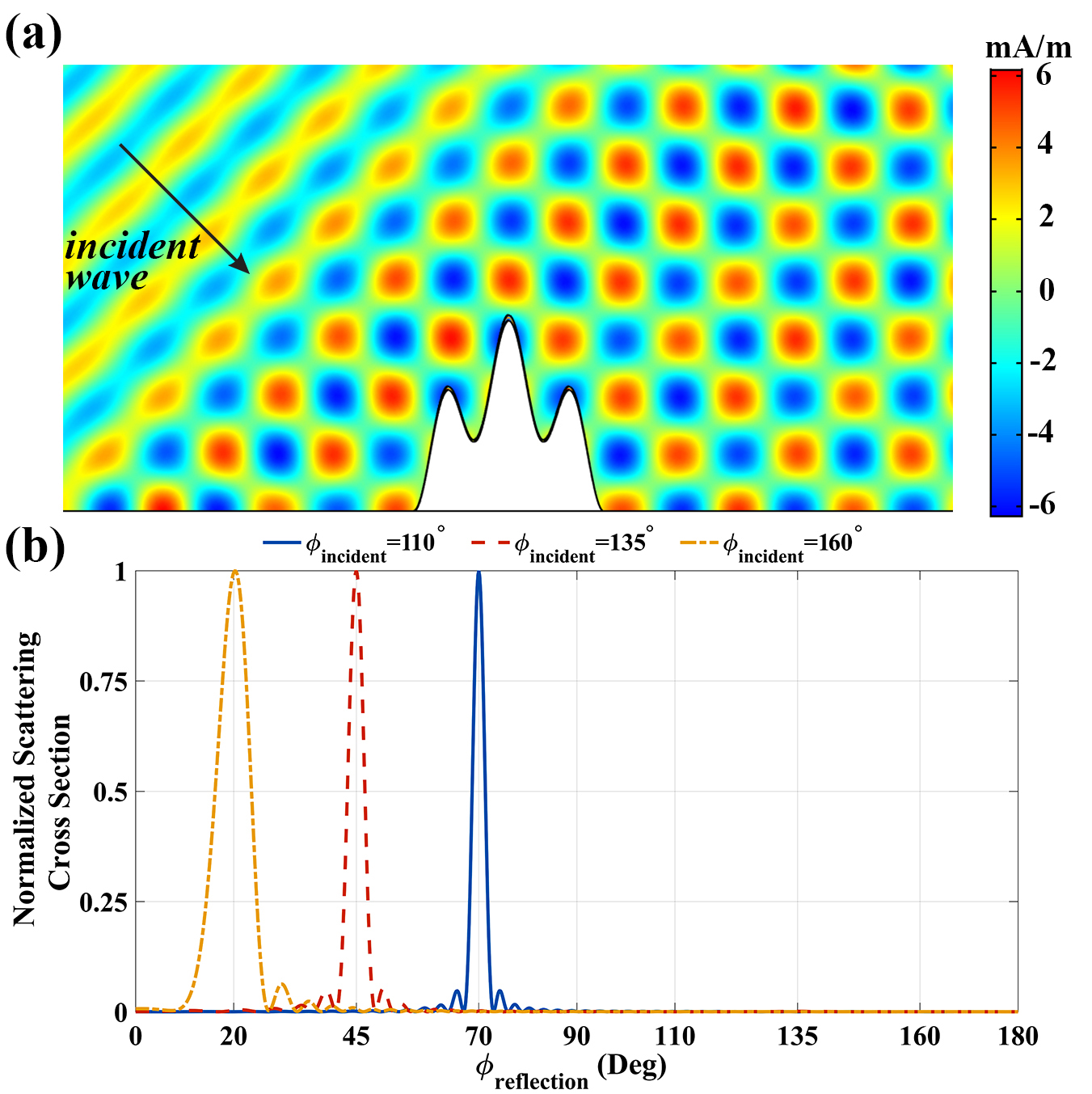}
	\caption{ Performance of the invisibility carpet cloak based on IAM for an arbitrary shape object with $\alpha=0.03$. (a) Magnetic field distribution at $\phi_{inc}=135^\circ$ . (b) Far-field patterns at $\phi_{inc}=110^\circ$, $\phi_{inc}=135^\circ$ and $\phi_{inc}=160^\circ$.}
	\label{fgr:fig3}
\end{figure}


 Although the propounded cloak can conceal the objects with ultrathin coating  layer from all viewing angles, this challenging question is raised that how  an IAM could be realized in practice. To answer this question, we will take the advantage of EMT  so as to validate the performance of an implemented triangle shape carpet cloak. With the assumption of TM  polarization, the only important parameters are $\varepsilon_{xx}=0$, $\varepsilon_{yy}={1+(f'(x))^2}/\alpha$ and $\mu_{zz}=1/\alpha$. As discussed above, when $\alpha \rightarrow 0$, the IAM with $\varepsilon=diag[0,\infty]$ and $\mu=\infty$ is acquired where $diag$[·] represents a diagonal matrix.
  \begin{figure}[h]
	\centering
	\includegraphics[width=0.7\linewidth]{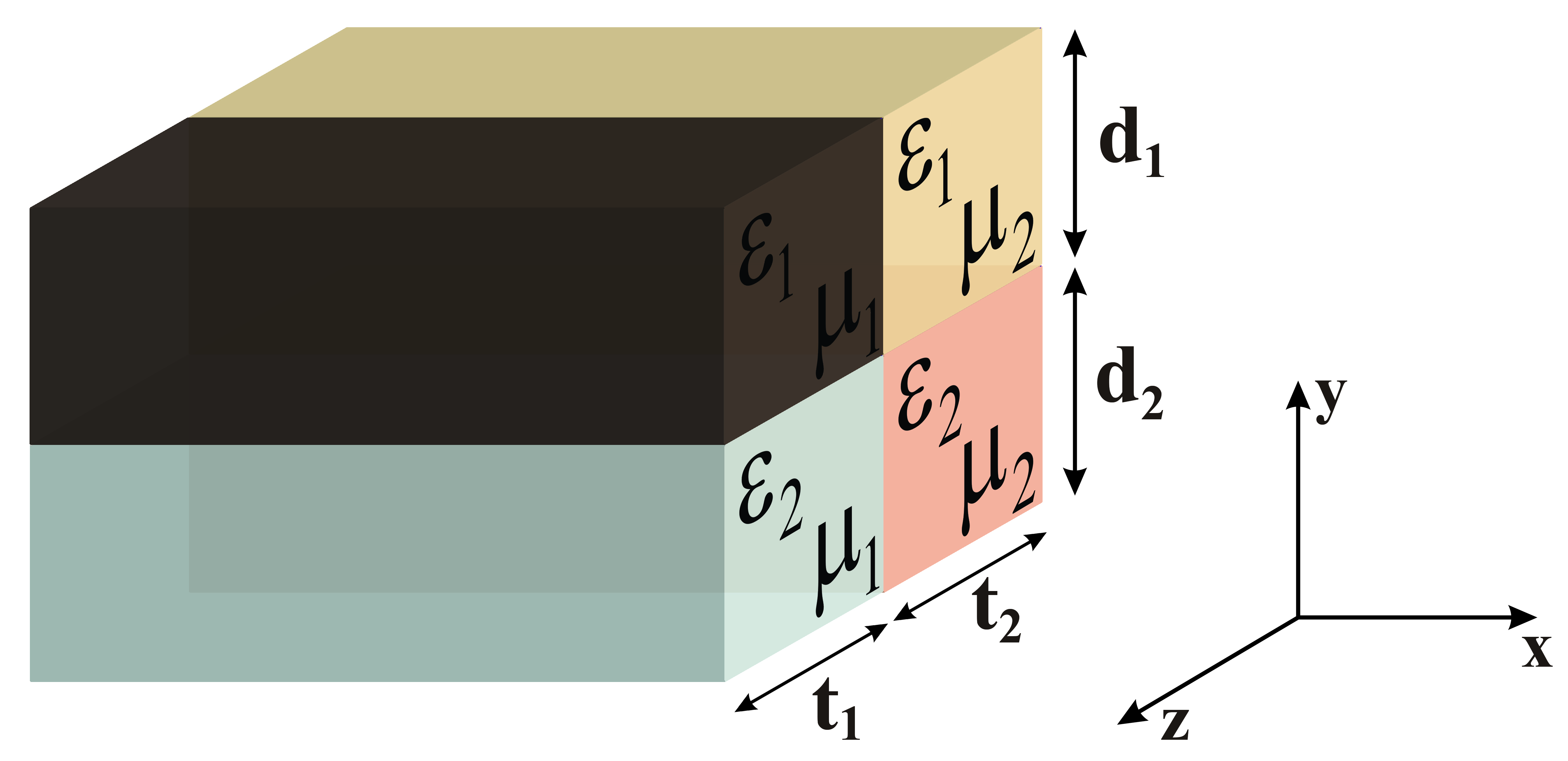}
	\caption{ A multi-layered structured that mimics an IAM. The electric permittivity and magnetic permeability are changed periodically in $y$ and $z$ directions, respectively.  }
	\label{fgr:fig4}
\end{figure}

\begin{figure}[h]
	\centering
	\includegraphics[width=\linewidth]{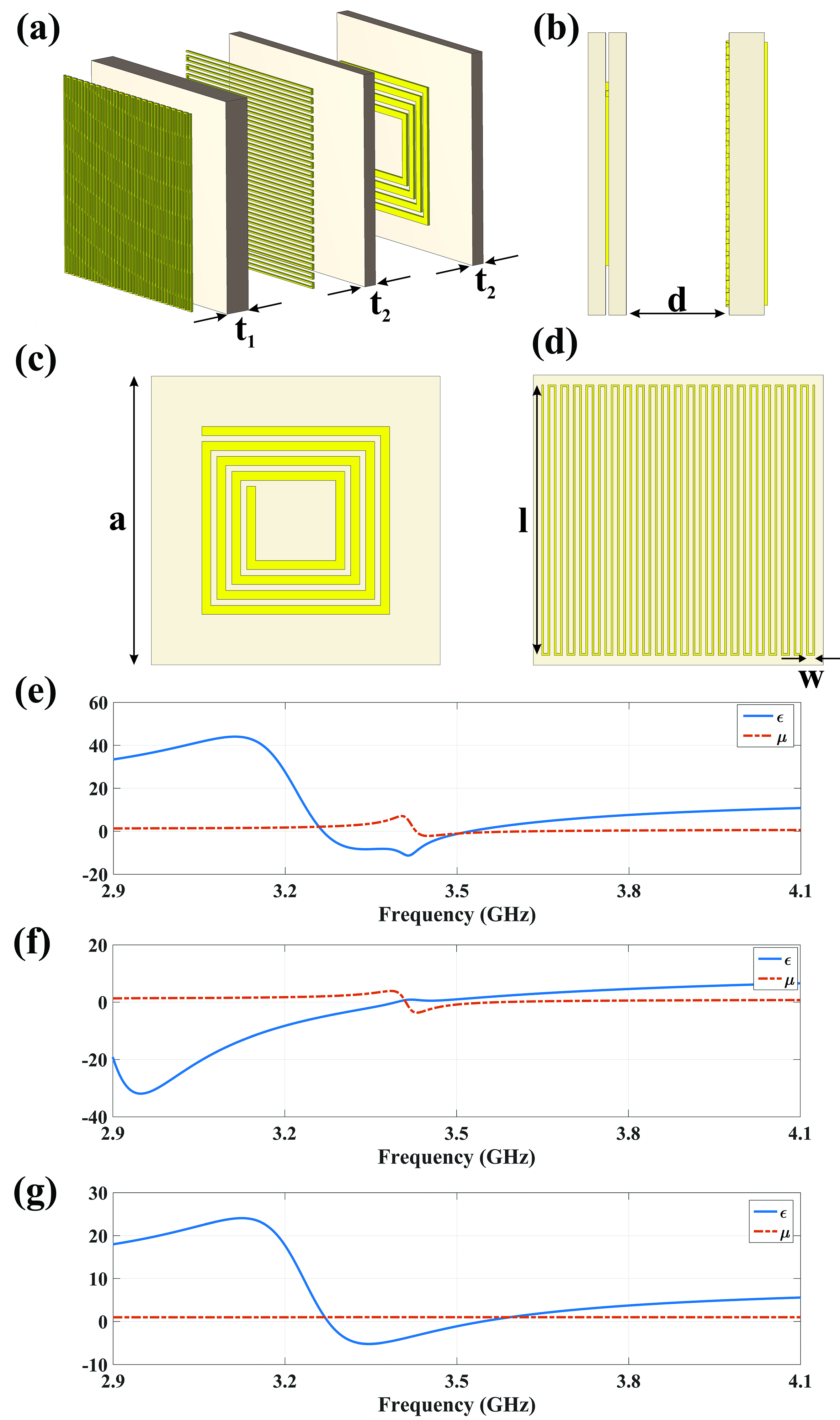}
	\caption{ (a) Composite of SR-meander line. (b) Side view of the proposed metamaterial. (c) The geometry of SR which controls the $\mu$ value. (d) The geometry of meander line which controls the $\varepsilon$ value. (e) The retrieved materials  for $\varepsilon=-1$, $\mu=-1$ with parameters $t_1=0.2 mm$, $t_2=0.1 mm$, $d=0.58 mm$, $a=1.7 mm$, $l=1.5 mm$, $w=0.045 mm$ and the SR has four turns, the inner radius of $0.226 mm$ and the gap value is $0.033 mm$. (f) The retrieved materials  for $\varepsilon=1$, $\mu=-1$ with parameters $t_1=0.2 mm$, $t_2=0.1 mm$, $d=0.58 mm$, $a=2 mm$, $l=1.49 mm$, $w=0.06 mm$ and the SR has four turns, the inner radius of $0.216 mm$ and the gap value is $0.053 mm$. (g) The retrieved materials  for $\varepsilon=-1$, $\mu=1$ with parameters $t_1=0.2 mm$,  $a=1.7 mm$, $l=1.54 mm$, $w=0.045 mm$.}
	\label{fgr:fig5}
\end{figure}
 \begin{figure}[h]
	\centering
	\includegraphics[width=\linewidth]{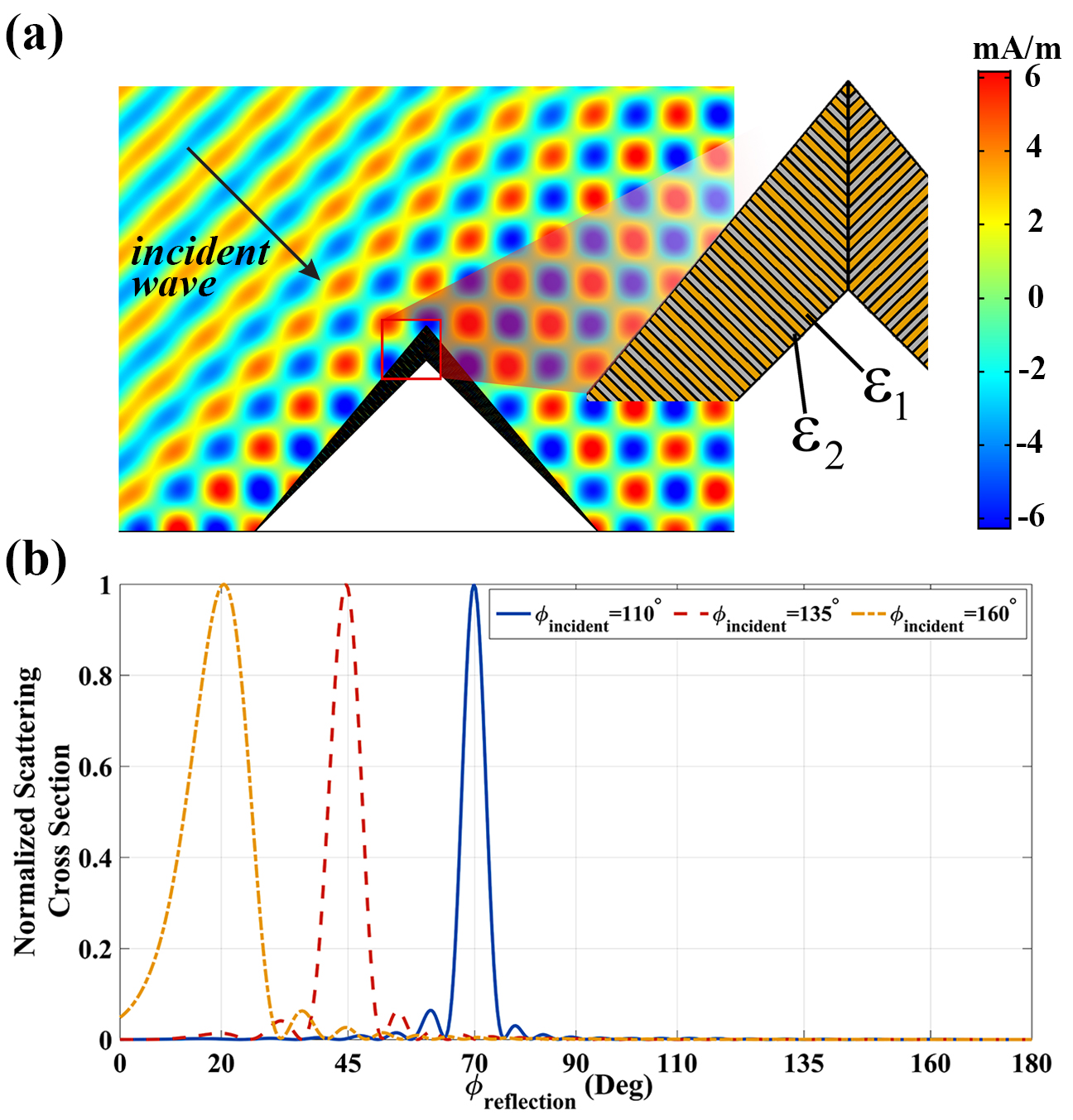}
	\caption{ Performance of the realized carpet cloak for a triangle shape object with $\alpha=0.2$. (a) Magnetic field distribution at $\phi_{inc}=135^\circ$ . (b) Far-field patterns at $\phi_{inc}=110^\circ$, $\phi_{inc}=135^\circ$ and $\phi_{inc}=160^\circ$.}
	\label{fgr:fig6}
\end{figure}
 According to EMT, the multi-layered structure displayed in Fig. 4 that consists of four isotropic blocks with constitutive parameters  $\varepsilon_{1,2}$ and $\mu_{1,2}$, can behave as an anisotropic medium with the parameters of IAM, if the thickness of each layer is  much smaller than the operative wavelength. In general, we have the following effective parameters:
 \begin{align}
 &\varepsilon_{xx}=f\varepsilon_1+(1-f)\varepsilon_2
 \\ \nonumber
 &\frac{1}{\varepsilon_{yy}}=\frac{f}{\varepsilon_1}+\frac{1-f}{\varepsilon_2}
 \\ \nonumber
 &\frac{1}{\mu_{zz}}=\frac{p}{\mu_1}+\frac{1-p}{\mu_2}
 \end{align}
in which $f=d_1/(d_1+d_2)$ and $p=t_1/(t_1+t_2)$. Based on Eq.(3), if we set $\varepsilon_{1,2}=\pm 1$, $\mu_{1,2}=\pm 1$, $f=0.5$ and $p=0.5$, then the desired IAM parameters can be attained, where its main axis is aligned toward $x$-direction. To construct each of these layers, a composite of SR-meander line array is utilized as depicted in Fig. 5(a). As proof of principle, a triangle shape carpet cloak with  $\alpha=0.2$ is implemented through the EMT. Since the $\alpha$ value is not too small, the required constitutive materials are obtained as $\varepsilon_{xx}=0.082$, $\varepsilon_{yy}=12.08$ and $\mu_{zz}=6$ after the diagonalization. Based on Eq.(3), one can simply implement this coating layer by using the introduced structure in Fig. 4 with the parameters $\varepsilon_{1}=1$, $\varepsilon_{2}=-1$, $\mu_{1}=1$, $\mu_{2}=-1$, $f=0.541$ and $p=0.583$. For any other shape of concealed objects and $\alpha$ values, the same materials $(\varepsilon_{1,2}=\pm 1,\mu_{1,2}=\pm 1 )$ could be used but with a different set of the $f$ and $p$ parameters. To achieve the required materials in each block in Fig. 4, the geometry of  SR-meander line unit cells are optimized, except the brown block which consists of air material or which is filled by $\varepsilon_1=1$ and $\mu_1=1$ material. The SRs are designed to have an adjustable control on $\mu_z$ in the cream and orange blocks, which are sandwiched between two RO3010 substrate ($\varepsilon_r=10.2, tg \delta=0.0022$) to reduce the coupling effect between the adjacent layers (see Fig. 5(b)). In addition, to obtain the same values of $\varepsilon_{xx}$ and $\varepsilon_{yy}$, two similar meander lines are printed in both top and bottom sides of the RO3010 substrate, with a $90^\circ$ rotation with respect to each other. The metamaterial building blocks are designed and simulated with CST Microwave Studio commercial software, where their retrieved material parameters are plotted in Figs. 5(e),  5(f) and 5(g) for the orange, cream and gray blocks, respectively. As can be seen,
from the retrieved parameters, the designed metamaterials are
capable of possessing the appropriate values at the desired frequency 3.5 GHz. \par
The triangle shape carpet cloak with  $\alpha=0.2$ is realized with the retrieved parameters of the propounded unit cell (see Fig. 6(a)). The metamaterial blocks are perpendicular to the object surface which is shown in the figure inset. As illustrated in Fig. 6(b), the designed metamaterials are capable of mimicking the IAM behavior and concealing the object from all viewing angles.
It is worth  mentioning that regardless of the object geometry, one can exploit the introduced IAM and achieve the cloaking functionality only with the four propounded building blocks $(\varepsilon_{1,2}=\pm 1,\mu_{1,2}=\pm 1 )$ perpendicular to the object surface and setting the $f(=d_1/(d_1+d_2))$ values.

 In conclusion, in this paper, we have presented a new material based on coordinate transformation capable of obviating the conventional challenges of carpet cloaks. The attained material which is called IAM, makes the carpet cloak extremely thin while the cloaking functionality remain unchanged for all viewing angles. Numerical simulations are proposed  corroborating well the validity and effectiveness of the propounded approach. In addition, the metamaterial building blocks that are competent to mimic the behavior of an IAM are proposed  and designed.  Therefore, when the geometry of the concealed object is changed, the same blocks could be used again with different proportions of the blocks thickness. This fact will, in turn, makes the proposed material as a good candidate to be used in real-life scenarios.  Then, as a proof of concept, a triangle  shape carpet cloak was realized with the aid of EMT and the designed metamaterials. It was observed that the simulation results exhibit good agreement with the theoretical predictions, which corroborates the correctness of the designed procedure. We believe that the newly proposed material in this paper constitutes a significant step towards the invisibility cloaks.


\begin{thebibliography}{1}
	
	\bibitem{Alu}
	Alù, Andrea, and Nader Engheta. "Achieving transparency with plasmonic and metamaterial coatings." Physical Review E 72, no. 1 (2005): 016623.
	
	\bibitem{Pendry}
	Pendry, John B., David Schurig, and David R. Smith. "Controlling electromagnetic fields." science 312, no. 5781 (2006): 1780-1782.
	
	\bibitem{kwon}
	Kwon, Do-Hoon, and Douglas H. Werner. "Polarization splitter and polarization rotator designs based on transformation optics." Optics Express 16, no. 23 (2008): 18731-18738.
	\bibitem{Abdolali}
	Abdolali, Ali, Hooman Barati Sedeh, and Mohammad Hosein Fakheri. "Geometry free materials enabled by transformation optics for enhancing the intensity of electromagnetic waves in an arbitrary domain." Journal of Applied Physics 127, no. 5 (2020): 054902.
	\bibitem{Fakheri}
	Fakheri, Mohammad Hosein, and Ali Abdolali. "Angularly Dispersionless Scattering Patterns for Impenetrable Surfaces: A Straightforward Design Based on Transformation Optics." IEEE Antennas and Propagation Magazine (2020).
	\bibitem{Lai}
	Lai, Yun, Jack Ng, HuanYang Chen, DeZhuan Han, JunJun Xiao, Zhao-Qing Zhang, and Che Ting Chan. "Illusion optics: the optical transformation of an object into another object." Physical review letters 102, no. 25 (2009): 253902.
	\bibitem{Li}
	Li, Jensen, and John B. Pendry. "Hiding under the carpet: a new strategy for cloaking." Physical review letters 101, no. 20 (2008): 203901.
	\bibitem{Zhang}
	Zhang, Baile, Tucker Chan, and Bae-Ian Wu. "Lateral shift makes a ground-plane cloak detectable." Physical review letters 104, no. 23 (2010): 233903.
	\bibitem{Yu}
	Yu, Nanfang, Patrice Genevet, Mikhail A. Kats, Francesco Aieta, Jean-Philippe Tetienne, Federico Capasso, and Zeno Gaburro. "Light propagation with phase discontinuities: generalized laws of reflection and refraction." science 334, no. 6054 (2011): 333-337.
	\bibitem{Behroozinia}
	Behroozinia, Sahar, Hamid Rajabalipanah, and Ali Abdolali. "Real-time terahertz wave channeling via multifunctional metagratings: a sparse array of all-graphene scatterers." Optics Letters 45, no. 4 (2020): 795-798.
	\bibitem{Rajabalipanah}
	Rajabalipanah, Hamid, Kasra Rouhi, Ali Abdolali, Shahid Iqbal, Lei Zhang, and Shuo Liu. "Real-time terahertz meta-cryptography using polarization-multiplexed graphene-based computer-generated holograms." Nanophotonics 9, no. 9 (2020): 2861-2877.
	\bibitem{Jing}
	Zhang, Jing, Zhong Lei Mei, Wan Ru Zhang, Fan Yang, and Tie Jun Cui. "An ultrathin directional carpet cloak based on generalized Snell's law." Applied Physics Letters 103, no. 15 (2013): 151115.
	\bibitem{Huang}
	Huang, Yijia, Mingbo Pu, Fei Zhang, Jun Luo, Xiong Li, Xiaoliang Ma, and Xiangang Luo. "Broadband functional metasurfaces: achieving nonlinear phase generation toward achromatic surface cloaking and lensing." Advanced Optical Materials 7, no. 7 (2019): 1801480.
	\bibitem{Jiang}
	Jiang, Zijie, Qingxuan Liang, Zhaohui Li, Tianning Chen, Dichen Li, and Yang Hao. "A 3D Carpet Cloak with Non‐Euclidean Metasurfaces." Advanced Optical Materials 8, no. 19 (2020): 2000827.
	\bibitem{Barati}
	Barati, H., M. H. Fakheri, and A. Abdolali. "Experimental demonstration of metamaterial-assisted antenna beam deflection through folded transformation optics." Journal of Optics 20, no. 8 (2018): 085101.
	
	\bibitem{Balanis}
	Balanis, Constantine A. Advanced engineering electromagnetics. John Wiley and Sons, 1999.

	
	
	
\end{thebibliography}
\end{document}